\documentclass[floats,showpacs,aps,twocolumn,prb]{revtex4}
\usepackage{graphicx,graphics,epsfig}
\usepackage{dcolumn}
\usepackage{bm,amsmath,amssymb}

\begin{document}

\title{Effects of polarization on the transmission and localization of
classical waves in weakly scattering metamaterials}

\author{Ara A. Asatryan$^1$, Lindsay C. Botten$^1$, Michael A. Byrne$^1$, Valentin D. Freilikher$^{2}$, Sergey A. Gredeskul$^{3,4}$, Ilya V. Shadrivov$^4 $, Ross C. McPhedran$^5$, and Yuri S. Kivshar$^4$}
\affiliation{$^1$Department of Mathematical Sciences, Centre for Ultrahigh-bandwidth Devices for Optical Systems (CUDOS), University of Technology, Sydney, NSW 2007, Australia\\
$^2$ Department of Physics, Bar-Ilan University, Raman-Gan, 52900, Israel\\
$^3$ Department of Physics, Ben Gurion University of the Negev, Beer Sheva, 84105, Israel\\
$^4$ Nonlinear Physics Center and CUDOS, Research School of Physics and Engineering, Australian National University, Canberra, ACT 0200, Australia\\
$^5$School of Physics and CUDOS, University of Sydney, Sydney, NSW 2006, Australia}

\begin{abstract}
We summarize the results of our comprehensive analytical and numerical studies of the effects of polarization on the Anderson localization of classical waves in one-dimensional random stacks. We consider homogeneous stacks composed entirely of normal materials or metamaterials, and also mixed stacks composed of alternating layers of a normal material and s metamaterial. We extend the theoretical study developed earlier for the case of normal incidence [A. A. Asatryan {\em et al}, Phys. Rev. B \textbf{81}, 075124 (2010)] to the case of off-axis incidence. For the general case where both the refractive indices and layer thicknesses are random, we obtain the long-wave and short-wave asymptotics of the localization length over a wide range of incidence angles (including the Brewster ``anomaly'' angle). At the Brewster angle, we show that the long-wave localization length is proportional to the square of the wavelength, as for the case of normal incidence, but with a proportionality coefficient substantially larger than that for normal incidence. In mixed stacks with only refractive-index disorder, we demonstrate that  {\em p}-polarized waves are strongly localized, while for {\em s}-polarization the localization is substantially suppressed, as in the case of normal incidence. In the case of only thickness disorder, we study also the transition from localization to delocalization at the Brewster angle.
\end{abstract}

\maketitle

\bigskip

%%%%%%%%%%%%%%%%%%%%%%%%%%%%%%%%%%%%%%%%%%%%%%%%%%

\section{Introduction}

\label{sec:intro}

Anderson localization is one of the most fundamental and fascinating phenomena in the physics of disordered systems. Predicted in the seminal paper of Anderson~\cite{Anderson} for spin excitations, and extended to the case of electrons and other one-particle excitations in solids (see, for example, Ref.~\cite{LGP}) and also applied to classical waves~\cite{John84}, this very general phenomenon has become a paradigm of modern physics \cite{Ping07}. Despite considerable efforts, the theoretical framework of Anderson localization in higher dimensions ($D>1$) is far from  complete \cite{Lage}, especially in the case of classical waves where the effects of absorption \cite{Frei2}, gain \cite{Bena,Asatryan}, and polarization \cite{PingP,Sipe,Bliokh} are significant.

In contrast, the one-dimensional case ($D=1)$ has been studied extensively for both quantum mechanical and classical waves (see, e.g., Refs.~\cite{LGP,Frei}). In the systems with short-range correlated disorder, it is known that all states are localized~\cite{Mott,Furst}. One of the main manifestations of localization is the exponential decay of the amplitude of a wave propagating through an infinite disordered sample. This decay is the result of the interference of multiply scattered waves, and its spatial rate is called the Lyapunov exponent, $\gamma$, whose inverse value $\gamma ^{-1}$ is a characteristic length describing localization in an infinite sample. By itself, however, the reciprocal of the Lyapunov exponent does not provide comprehensive information about the transport properties of disordered media for all cases (see Ref.~\cite{Asatryan10} for details). Moreover, it is unlikely that this quantity can be measured directly, at least in the optical regime.

A further manifestation of localization is the exponential decay of the transmission coefficient of a long, finite sample. The characteristic length of this decay is the transmission length $l_{T}$, also denoted by $l_N$ in the case of a sample of $N$ layers. In the localized regime, where this length is much smaller than the sample size, in general the transmission length coincides with the localization length $l$, and also with the inverse Lyapunov exponent $\gamma ^{-1}$ .

The Anderson localization of classical waves in one-dimensional disordered systems has been studied in detail \cite{Asatryan,PingP,Varuz,deSterke,Luna}) and it has been shown that in the long wave region where the interference is weak, the localization length demonstrates a universal behavior, growing in proportion to the square of the wavelength, i.e., $l \propto \lambda ^{2}$.

In recent years, we have witnessed the rapid emergence of a new field of research in metamaterials---artificial materials which exhibit a negative refractive index \cite{Veselago,Pendry,Bliokh04,Sok}. In such materials, the wave vector $\mathbf{k}$, the electric field vector $\mathbf{E}$ and the magnetic field vector $\mathbf{H}$ form a left-handed (\emph{LH}) coordinate system, in contrast to the right-handed system (\emph{RH}) that is applicable to normal or regular materials; for this reason, metamaterials are sometimes referred to as left-handed materials. In metamaterials, the directions of the phase velocity and the energy flow are opposite. This feature can strongly affect Anderson localization in metamaterials. Indeed, in stratified media formed of alternating layers of normal materials and metamaterials, the phase accumulated during propagation through a right-handed layer diminishes in its propagation in a left-handed layer \cite{Asatryan07} and, as a consequence, interference will be weakened and localization suppressed.

Already, one of the first study\cite{Asatryan07} of Anderson localization in the presence of metamaterials has  revealed the striking behavior that, in the particular case of an alternating  stack of right- and left-handed layers of the same thickness and randomly varying refractive indices, the localization is strongly suppressed. Its functional form in the long-wave region changes from the standard behavior of $l \propto \lambda^{2}$ to $l \propto \lambda ^{6}$ and, subsequently, it was shown~\cite{Asatryan10} that in such stacks the localization length differs from the inverse of the Lyapunov exponent---the first such example of this surprising behavior.

While the disorder is one-dimensional, the random stacks are actually three-dimensional objects and so the vector nature of the propagating field and, in particular, its polarization can strongly influence localization, leading to its suppression or even complete delocalization.

For stacks comprising only normal layers, the effects of polarization have been extensively studied. The problem  has been studied using an approach based on stochastic differential equations\cite{PingP}, with excellent agreement obtained at short wavelengths (for normal incidence), and also in the long wavelength limit, between the theoretically predicted localization length and that obtained by direct simulation. The delocalization associated with the Brewster angle anomaly has been studied in the long wave limit using an effective medium approach \cite{Sipe}, and at short wavelengths in the framework of a random phase approximation \cite{Bliokh}. Numerical simulations for the off-axis case are given in Ref. \cite{Malish} while, in Ref. \cite{Xu}, the polarization properties of localization have been considered experimentally.

Although the approaches described in Refs \cite{Asatryan,PingP,Sipe,Bliokh,Malish,Xu} give rich information about the effects of polarization on the properties of localization in normal materials, a general expression for the localization length, applicable for broad range of input parameters (including at the Brewster anomaly angle) is still missing. Furthermore, there are no results available for the effects of polarization on localization in the presence of metamaterials.

In this paper, we extend our earlier study~\cite{Asatryan10} to the case of off-axis incidence and provide a comprehensive analytical and numerical treatment of the effects of polarization on the localization length $l$.  We consider both homogeneous stacks, formed by only normal material layers or by only metamaterial layers, and also mixed stacks formed by an alternating sequence of right- and left-handed layers with random thicknesses and refractive indices. We derive explicit asymptotic expressions for the localization length at the short and long wavelength limits that are applicable for the Brewster angle, and demonstrate their excellent agreement with direct simulation.

The paper is organized as follows. In Sec.~\ref{sec:theor}, we describe the model and outline the theoretical treatment. The derivation of the asymptotic forms for the localization length at short and long wavelengths is presented in Sec.~\ref{asymp}. In Sec.~\ref{numerics}, we present the results of numerical simulations for the localization length for \emph{s}- and \emph{p}- polarizations in both homogeneous and mixed stacks. Here, we adopt the conventional definition for polarization, with \emph{s}- and \emph{p}-polarization referring respectively to the cases in which the electric and magnetic fields are perpendicular to the plane of incidence. Finally, in Sec.~\ref{theta}, the dependence of the localization length on the angle of incidence, at a fixed wavelength, is considered.

\section{Theoretical studies}
\label{sec:theor}

\subsection{Model}
\label{subsec:model}

\begin{figure}[h]
\center{\scalebox{0.25}{\includegraphics{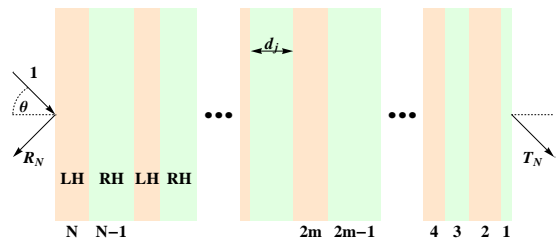}}}
\caption{(Color on line) The geometry of the structure. }
\label{geom}
\end{figure}

We consider the transmission and localization properties of a one-dimensional, multi-layered, disordered stack which consists of $N$ layers composed of either right-handed or normal ($r$) materials, left-handed ($l$) metamaterials, or mixed stacks comprising alternating layers ($r$ and $l$) of each (see Fig.~\ref{geom}).

%%%%%%%%%%%%%%%%%%%%%%%%%%%%%%%%%%%%%%%%%%%%%%%%%%%%%%%%%%%%%%%%%%%%%%%%
%%%%%%%%%%%%%%%%%%%%%%%%%%%%%%%%%%%%%%%%%%%%%%%%%%%%%%%%%%%%%%%%%%%%%%%%

All layers are statistically independent and, in the most general case, the thickness of each layer, and its dielectric and magnetic permittivities, are random quantities with given probability densities. All lengths in the problem are measured in the unit of the mean layer thickness and, therefore, are dimensionless quantities.

The main subject of our interest is the transmission length \cite{LGP,Ping07}
\begin{equation}
l_N =- \frac{N}{\langle \ln |T_{N}| \rangle},  \label{TL}
\end{equation}
where $T_N$ is the transmission coefficient of the $N$ layer stack for a plane wave with a given incidence angle (relative to the surface normal of the stack). Angular brackets are used to denote averaging over realizations of all random parameters. In the limit of a stack of infinite length, i.e., $N\to\infty$, the transmission length coincides with the localization length,
i.e.,
\begin{equation}
l= \lim_{N\to\infty} l_N.  \label{LL}
\end{equation}

The calculation of the transmission length (\ref{TL}) requires the transmission coefficient $T_N$ of the $N$-layer stack for a wave of a given polarization and a given incidence angle. Such a calculation can be based on the transfer matrix method \cite{Varuz}, the interface iteration method \cite{Asatryan}, and the layer iteration method \cite{Asatryan07}. We choose the last of these and build on the treatment that was used successfully in our previous study \cite{Asatryan10} for the case of normal incidence.

The method is based on the \emph{exact} iteration of the recurrence relations
\begin{eqnarray}
T_{n} = \frac{T_{n-1} t_{n}}{1-R_{n-1} r_{n}},  \label{recT} \\
R_{n} = r_{n} + \frac{R_{n-1} t^2_{n}}{1-R_{n-1}r_{n}}  \label{rec}
\end{eqnarray}
for the total transmission $T_{n}$ and total reflection $R_{n}$ coefficients of a $n$-layer stack, $n=1, \ldots, N$, in which both the input and output media are free space, and with initial conditions set to $T_0=1$ and $R_0=0$. Here $t_{n}$ and $r_{n}$ are the transmission and reflection coefficients of the $n^{\mathrm{th}}$ layer, with layers being enumerated from $n=1$ at the rear of the stack through to $n=N$ at the front.

Such an approach is quite general and is applicable to an arbitrary choice of polarization (\emph{s}) or (\emph{p}), the angle of incidence $\theta$, the type of layer ($r$ or $l$), the wavelength $\lambda$, and the amount of absorption or gain. All these input data are incorporated into the transmission and reflection characteristics $t_{n}$ and $r_{n}$ of a single layer.

The recurrence relations (\ref{recT}) and (\ref{rec}), together with the definitions (\ref{TL}) and (\ref{LL}) enable the numerical calculation of the transmission length $l_N$ in the most general case. If the calculated transmission length $l_N$ is much smaller than the stack length $N$, and is independent of $N$, then $l_N$ may be identified as the localization length, i.e., $l_N=l$.

In the work reported in this paper, we deal with a specific model in which the dimensionless thicknesses $d$ of each layer are independent, identically distributed, random variables, $d=1+\delta_d$, where $\delta_d$ is uniformly distributed in the range $[-Q_d,Q_d]$ with $0 \leq Q_d < 1.$ The dielectric and magnetic permittivities of the layers are represented in the form
\begin{equation}
\varepsilon=\pm(1+\delta_{\nu})^2, \ \ \ \mu=\pm 1,  \label{epsi}
\end{equation}
where the upper and lower signs respectively correspond to a normal material or a metamaterial. The random part $\delta_\nu$ of the refractive index,
\begin{equation}
\nu=\pm(1+\delta_\nu),  \label{nref}
\end{equation}
is uniformly distributed in the range $[-Q_\nu,Q_\nu]$, $0\leq Q_\nu<1.$ Accordingly, the model takes into account both refractive index disorder and layer thickness disorder.

\subsection{Theoretical analysis}
\label{analytics}
Our theoretical treatment is based on a weak scattering approximation (WSA) in which the magnitude of the reflection coefficient of each layer $|r_{n}|\ll 1$ is the primary small parameter in the theory. To understand when this condition is valid and may be used, we next consider explicit expressions for the reflection $r$ and transmission $t$ coefficients of any
single layer:
\begin{eqnarray}
r = \frac{\rho(1-e^{2i\beta})}{1-\rho^2e^{2i\beta}}, \qquad
t = \frac{(1-\rho^2)e^{i\beta}}{1-\rho^2e^{2i\beta}},  \label{t}
\end{eqnarray}
where $\rho$ is Fresnel interface reflection coefficient given by
\begin{eqnarray}  \label{ro}
\rho = \frac{\mathcal{Z}\cos\theta_{\nu}-\cos\theta} {\mathcal{Z}%
\cos\theta_{\nu}+\cos\theta}, \ \ \ \ \mathcal{Z}=\left\{
\begin{array}{ccc}
Z^{-1}, &  & s-\text{polarization} \\
Z, &  & p-\text{polarization}. \\
&  &
\end{array}
\right.
\end{eqnarray}
In these equations,
\begin{equation}  \label{notations}
\beta=k d \nu \cos\theta_{\nu}, \ \ \ \cos\theta_{\nu}=\sqrt{1-\frac{%
\sin^{2}\theta}{\nu^{2}}}, \ \ \ k=2\pi/\lambda,
\end{equation}
$\lambda$ is the dimensionless free space wavelength, and
\begin{equation}
Z= \sqrt{\frac{\mu}{\varepsilon}}= \frac{1}{1+\delta_{\nu}}>0  \label{zlab}
\end{equation}
is the layer impedance relative to the background (free space).

There are two cases in which the magnitude of the single layer reflection coefficient is small, i.e., $|r|\ll 1$. The first is characterized by weak refractive index disorder $Q_{\nu}\ll 1$ and corresponds to the incidence angle $\theta$ being smaller than its critical value
\begin{equation}
\label{theta-c}
\theta_{c}= \sin^{-1}(1-Q_{\nu}).
\end{equation}
In this case it is the small magnitude of the Fresnel reflection coefficient $|\rho|\ll 1$ which leads to a small, single layer reflection coefficient ($|r|\ll 1$) at all wavelengths. The second case corresponds to the long wavelength limit ($\lambda\gg 1$) in which the single layer reflection coefficient is small ($|r|\ll 1$) for an arbitrary incidence angle $\theta$ due to the asymptotically small value of the multiplier $|1-e^{2i\beta}| \propto \beta \ll 1$ which appears in the expression for $r$ in Eq. (\ref{t}).

Within the WSA approximation, we commence the derivation of general forms
with the linearized recurrence relations (\ref{rec})
\begin{eqnarray}  \label{rec3}
\ln T_{n} & = & \ln T_{1,n-1} + \ln t_{n}+R_{n-1}r_{n},  \notag \\
R_{n} & = & r_{n}+ R_{n-1} t_n^2,
\end{eqnarray}
and solve them to yield
\begin{equation}  \label{lnT1}
\ln T_{N}=\sum_{j=1}^{N}\ln
t_{j}+\sum_{m=2}^{N}\sum_{j=m}^{N}r_{j-m+1}r_{j}%
\prod_{p=j-m+2}^{j-1}t_{p}^{2},
\end{equation}
from which we may compute ensemble averages.

In what follows, we summarize the key theoretical results \cite{Asatryan10} applicable to mixed and homogeneous stacks.
%%%%%%%%%%%%%%%%%%%%%%%%%%%%%%%%%%%%%%

\subsubsection{Mixed stacks}
\label{subsubsec:mixed}

A mixed stack, which hereafter is abbreviated by M-stack, is composed of alternating layers of right-handed ($r$) and left-handed ($l$) materials (see Fig.~\ref{geom}). Within the model, there are simple relations that may be derived \cite{Asatryan10} between the averaged values of an analytic function $g$ of the transmission and reflection characteristics of single layers of left- and right-handed materials,
\begin{equation}  \label{symmetry}
\langle g(t_{r})\rangle =\langle g(t_{l})\rangle ^{\ast }, \qquad \langle
g(r_{r})\rangle =\langle g(r_{l})\rangle ^{\ast }.
\end{equation}

As a consequence, the transmission length of a M-stack depends only on the properties of a single right-handed layer and is expressible in terms of only three averaged quantities: $\langle r\rangle^{2} $, $\langle \ln t\rangle $, and $\langle t^{2}\rangle $, in which the subscript $r$ (referring to a right-handed layer) has been omitted:
\begin{equation}  \label{PRB1}
\frac{1}{l_N}=\frac{1}{l}+\left(\frac{1}{b}-\frac{1}{l}\right) f\left( \frac{%
N}{\bar{l}_{m}}\right).
\end{equation}
Here,
\begin{equation}  \label{PRB2}
\frac{1}{l}=-\mathrm{Re} \ \langle \ln t\rangle -\frac{|\langle r\rangle
^{2}|+\mathrm{Re}\left( \langle r\rangle^{2}\langle t^{2}\rangle ^{\ast
}\right) }{1-|\langle t^{2}\rangle |^{2}},
\end{equation}%
is the inverse localization length,
\begin{eqnarray}  \label{PRB3}
\frac{1}{b} &=&\frac{1}{l}-\frac{2/\bar{l}_{m}}{1-\exp (-2/\bar{l}_{m})}%
\times  \notag \\
&&\left( \frac{|\langle r\rangle^{2} |+\mathrm{Re}\left( \langle r\rangle
^{2}\langle t^{2}\rangle ^{\ast }\right) }{1-|\langle t^{2}\rangle |^{2}}-%
\frac{|\langle r\rangle^{2} |}{2}\right)
\end{eqnarray}%
is the inverse ballistic length, and
\begin{equation}  \label{f}
f(z)=\frac{1-\exp (-z)}{z}.
\end{equation}

To characterize the transition from localization to ballistic propagation, we introduce the \emph{crossover length} for a M-stack \cite{Asatryan10},
\begin{equation}  \label{l}
\bar{l}_m=-\frac{1}{\ln |\langle t^{ 2}\rangle |},
\end{equation}
and two characteristic wavelengths $\lambda_1$ and $\lambda_2$ defined by
\begin{equation}  \label{lam12}
N=l\left(\lambda_1(N)\right),\,\,\,N=\overline{l}\left(\lambda_2(N)\right).
\end{equation}
For long wavelengths, that part of the spectrum for which $\lambda\ll\lambda_1(N)$ corresponds to the localization regime, for which $l_N=l$. In turn, the wavelength range  $\lambda\gg\lambda_2(N)$ corresponds to ballistic propagation and $l_N=b$. The region $\lambda_1<\lambda<\lambda_2$ is the transition region from localization to ballistic propagation. In what follows, we characterize these regions with long wavelength asymptotes for the localization length $l$ and the crossover length $\overline{l}$.
%%%%%%%%%%%%%%%%%%%%%%%%%%%%%%%%%%%%

\subsubsection{Homogeneous stacks}
\label{subsubsec:homo}

The homogeneous stack (abbreviated from here as a H-stack) is composed of
layers of the same type of material (either $r$ or $l$). By virtue of the
symmetry relations (\ref{symmetry}), the results for statistically
equivalent stacks of either normal materials or metamaterials are identical.
%For definiteness,  we choose normal (right-handed) layers.

The transmission length of a H-stack is then
\begin{equation}  \label{FinalN1}
\frac{1}{l_N} = \frac{1}{l}+\frac{1}{N} \ \mathrm{Re }\left[\langle r
\rangle^2 \frac{1-\langle t^2\rangle^N}{(1-\langle t^2\rangle)^2}\right],
\end{equation}
while the inverse localization length is given by
\begin{equation}  \label{FinalN2}
\frac{1}{l}= -\mathrm{Re} \ \langle \ln t \rangle - \mathrm{Re }\frac{%
\langle r \rangle^2 }{1-\langle t^2\rangle}.
\end{equation}
It follows from Eq. (\ref{FinalN1}) that the crossover length $\overline{l}_{h}$ of the H-stack is
\begin{equation}  \label{crossL}
\overline{l}_{h}=-\frac{1}{|\ln\langle t^2 \rangle|},
\end{equation}
characterizing the transition from the near ballistic regime to the far ballistic regime of propagation.

\section{Asymptotic behavior of the transmission and localization lengths}

\label{asymp}

The results (\ref{PRB1}) - (\ref{PRB3}), (\ref{FinalN1}), (\ref{FinalN2}) for the transmission, localization and ballistic lengths of homogeneous and mixed stacks are quite general. In this Section we apply them to the specific model described in Subsection \ref{subsec:model}. We emphasize that within the model both two types of disorder are present and are taken into account in all intermediate calculations. However, in all final results we keep only the leading terms. The higher order corrections with respect to the relative perturbations $Q_{\nu,d}$ of the refractive index and thickness distributions are generally omitted.

\subsection{Short-wave asymptotics of the transmission length}
\label{subsec:short}

When the incidence angle $\theta$ is smaller than the critical angle $\theta_{c}$ (\ref{theta-c}), the short wavelength asymptotics of the localization length can be easily obtained from Eqs (\ref{PRB2}) and (\ref{FinalN2}). At short wavelengths, the phase of the field is a strongly fluctuating, random quantity so that $\langle t^2\rangle \approx 0$, $\langle r\rangle \approx \langle \rho\rangle$, and $\langle\ln |t|\rangle \approx \langle\ln(1-\rho^2)\rangle$. As a consequence, the localization length for both mixed and homogeneous stacks takes the form
\begin{equation}
\frac{1}{l}= -\langle \ln(1-\rho^2)\rangle - \langle \rho\rangle^{2}.
\label{ShAs3}
\end{equation}

For \emph{s}-polarization, the logarithmic term in Eq.~(\ref{ShAs3}) always dominates and so
\begin{equation}
\frac{1}{l}\approx \frac{Q_{\nu}^2}{12\cos^4\theta},  \label{shLasEa}
\end{equation}
while the second term in (\ref{ShAs3}) provides a higher order correction of order $O\left( Q_{\nu}^{4} \right)$. For \emph{p}-polarization, however, the corresponding correction cannot be omitted since the first term vanishes at the Brewster angle $\theta=\pi/4$. Thus, for \emph{p}-polarization,
\begin{eqnarray}
\frac{1}{l} \approx \frac{Q_{\nu}^2\cos^2(2\theta)}{12\cos^4\theta}+\frac{%
Q_{\nu}^4}{120\cos^8\theta}\times  \notag \\
\left ( \frac{569}{96} -8\cos2\theta + \frac{43}{8}\cos4\theta + \cos6\theta
+\frac{11}{96}\cos8\theta \right ) .  \label{shLasEaa}
\end{eqnarray}
Accordingly, at the Brewster angle, the localization length is given by
\begin{equation}
{l}=\frac{45}{4Q^4_{\nu}}.  \label{ShAs5}
\end{equation}
%and essentially exceeds that out of the Brewster angle which is $\propto %Q_{\nu}^{-2}$.
Note that in the case of normal incidence, the results (\ref{shLasEa}) and (\ref{shLasEaa}) coincide with those presented in Refs \cite{Asatryan,PingP,Varuz,deSterke}.

If the incidence angle $\theta$ exceeds the critical angle $\theta_{c}$ (\ref{theta-c}), then total internal reflection occurs (i.e., the magnitude of the Fresnel reflection coefficient becomes unity) and so the WSA fails in the short wave region. If the incident angle is sufficiently far above the critical value $\theta_{c}$, then the exponent $2i\beta$ in Eq. (\ref{t}) is real and negative and thus the magnitude of the single layer transmission coefficient is exponentially small. As a result, we obtain the following expression for the transmission length
\begin{equation}  \label{ShAs2}
\frac{1}{l_{N}} = \mathrm{Im }\langle \beta \rangle =k \mathrm{Im} \langle d
\sqrt{\sin^{2}\theta-\nu^{2}}\rangle, \quad \mathrm{for ~~}
\sin\theta>1-Q_{\nu}.
\end{equation}
The right hand side of this equation is independent of the stack length $N$ and so it formally coincides with the inverse localization length $l$. However, its origin is related to attenuation by tunneling, rather than to Anderson localization. In the attenuation regime, the transmission length does not distinguish the left- and right-handed layers since it depends on
the square of the refractive index $\nu$ and so is the same for equivalent $l$ or $r$ layers. Moreover, it does not distinguish the polarization of the light.

The average of Eq. (\ref{ShAs2}) can be calculated in closed form for the uniform distribution of the refractive index and we obtain the short wave asymptotic of the transmission length in the attenuation regime from the expression
\begin{eqnarray}
\frac{1}{l_{N}} & = & \frac{k}{4Q_{\nu}} \left[\left(\frac{\pi}{2}-\sin^{-1}
\frac{1-Q_{\nu}}{\sin\theta}\right)\sin^2\theta\right.  \notag \\
&-&\left. (1-Q_{\nu})\sqrt{\sin^2\theta-(1-Q_{\nu})^2}\right],  \label{shcl}
\end{eqnarray}
and see that it is proportional to the wavelength. We see also that it holds for both polarizations, and also for both H- and M-stacks.

The main contribution to this length in the short wave region coincides with that of the first terms in Eqs (\ref{PRB2}) and (\ref{FinalN2}) for the localization lengths of H- and M-stacks. These terms in Eqs (\ref{PRB2}) and (\ref{FinalN2}) dominate in the short wave region, as we will demonstrate below, and thus these terms give the correct values for the transmission length in the attenuation regime (for both short wave and long wave regions) despite the applicability
of the WSA being violated.

\subsection{Long-wave asymptotics for homogeneous stacks}
\label{subsec:long-homo}

\subsubsection{Homogeneous stacks: $s$-polarization}
\label{subsubsec:NLongE}

For long wavelengths, the transmission length can be deduced from the general result in Eq. (\ref{FinalN2}). In this limit, the mean values of $\langle \ln t\rangle$, $\langle t^2 \rangle $ and $\langle r \rangle $ that enter (\ref{FinalN2}) for \emph{s}-polarization take the form
\begin{eqnarray}
\langle \ln t \rangle & = & ik\cos\theta+\frac{ikQ^2_{\nu}}{6\cos\theta}-
\frac{k^2Q^2_{\nu}}{6\cos^2\theta}\left(1+\frac{Q_d^2}{3}\right)\notag,\\
& - & \frac{k^2Q^4_{\nu}}{40\cos^2\theta}\left (1+\frac{Q^2_d}{3}\right) ,
\label{1E} \\
\langle r \rangle  & =  &\frac{ikQ^2_{\nu}}{6\cos\theta}-\frac{k^2Q^2_{\nu}}{6\cos^2\theta}(2+\cos^2\theta +\frac{3Q^2_{\nu}}{10})\notag\\
& \times & (1+\frac{Q^2_d}{3})-\frac{ik^3Q^2_{\nu}}{9\cos\theta}(4+\cos^2\theta)\left (1+Q^2_d \right )\notag\\
& - & \frac{ik^3Q^4_{\nu}}{2\cos^3\theta}\left (\frac{2}{3}+\frac{\cos2\theta}{15}+\frac{Q^2_{\nu}}{28}\right )(1+Q^2_d)\notag\\
& - & \frac{ik^3Q^6_{\nu}}{56\cos^3\theta}(1+Q^2_d),
\label{2E} \\
\langle t^2 \rangle  & = & 1+2ik\cos\theta+\frac{ikQ^2_{\nu}}{3\cos\theta}-2k^2\cos^2\theta\notag\\
& - & \frac{2k^2Q_d^2}{3}\cos^2\theta-\frac{k^2Q^2_{\nu}}{3\cos^2\theta}(3+2\cos^2\theta)\notag\\
& \times & \left(1+\frac{Q^2_d}{3}\right)-
\frac{3k^2Q^4_{\nu}}{20\cos^2\theta}\left(1+\frac{Q^2_d}{3}\right).
 \label{3E}
\end{eqnarray}

Applying the results of Eqs (\ref{1E})-(\ref{3E}) in the expression for the
transmission length (\ref{FinalN1}), we obtain
\begin{eqnarray}
\frac{1}{l_N} & = & \frac{k^2Q_\nu^2}{6\cos^2\theta} \left (1+\frac{Q^2_{\nu}%
}{15}-\frac{Q^2_{\nu}}{3\cos^2\theta}\right )\left (1+\frac{Q_d^2}{3}\right)
\notag \\
& + & \frac{Q_\nu^4}{72\cos^4\theta}\left(\frac{\sin^2(kN\cos\theta)}{N}%
\right),  \label{ltrE0}
\end{eqnarray}
which is correct to the order of $Q^4_{\nu}$. This expression can be further simplified given that $Q_{\nu}\ll 1$ and $Q_d\ll 1$. In this approximation, the transmission length takes form
\begin{equation}
\frac{1}{l_N}=\frac{k^2Q_\nu^2}{6\cos^2\theta}\left [1+ \frac{NQ_\nu^2}{12}%
\left(\frac{\sin(kN\cos\theta)}{kN\cos\theta}\right)^2\right],  \label{ltrE}
\end{equation}
from which it follows that the localization length is given by
\begin{equation}
l= \frac{3\lambda^2\cos^2\theta}{2\pi^2Q_{\nu}^2}, \ \ \ l\leq N.
\label{LoLE}
\end{equation}
From this, it is seen that the localization length of a homogeneous stack  in the long wavelength limit does not depend on the thickness disorder for {\em s} polarization. We verified this through exact numerical calculations (see Fig.\ref{Fig2}).

Eq. (\ref{LoLE}) reproduces the result first derived in Ref. \cite{PingP} and, for normal incidence, coincides with the localization length derived more recently in  \cite{Asatryan10,Izra10}.  For wavelengths such that $l\geq N\geq \lambda/\cos\theta$, there is a near ballistic regime where the ballistic length $b$ is the same as the localization length
\begin{equation}
b= \frac{3\lambda^2\cos^2\theta}{2\pi^2Q_{\nu}^2}, \ \ \ l\geq N\geq %
\displaystyle{\frac{\lambda}{\cos\theta}}.  \label{balE}
\end{equation}

The crossover length in the far ballistic region (\ref{crossL}) is
\begin{equation}  \label{balHomSpol}
\overline{l}_{h}=\frac{\lambda}{4\pi\cos\theta},
\end{equation}
while the ballistic length $b_{f}$ in this region is
\begin{equation}
{b_f}=\frac{3\lambda^2\cos^2\theta}{2\pi^2Q_{\nu}^2}\left [1+ \frac{NQ_\nu^2%
}{12}\right]^{-1}.  \label{ltrLE}
\end{equation}

From Eqs (\ref{LoLE}), (\ref{balHomSpol}) and (\ref{lam12}), it follows that
\begin{eqnarray}
\lambda_1 &= & \frac{\pi Q_{\nu}}{\cos\theta}\sqrt{\frac{2N}{3}},
\label{L1HS} \\
\lambda_2 &=& 4\pi N \cos\theta,  \label{lam12HomS}
\end{eqnarray}
for all reasonable values of the input parameters such that $\lambda_1<\lambda_2$. For $\lambda<\lambda_1$, waves are localized with the localization length given by Eq. (\ref{LoLE}). The wavelength range $\lambda_1<\lambda<\lambda_2$ corresponds to the near ballistic regime in which the ballistic length $b$ is given by Eq. (\ref{balE}). The longer wavelengths, $\lambda\geq \lambda_2 $, correspond to the far ballistic regime in which the ballistic length $b_{f}$ is given by Eq. (\ref{ltrLE}).

\subsubsection{Homogeneous stacks: $\mathit{p}$-polarization}
\label{subsubsec:NLongH}

In the case of $p-$polarized waves, the mean values of $\langle \ln t\rangle$, $\langle t^2 \rangle $ and $\langle r \rangle^2 $, at longer wavelengths, take the form
\begin{eqnarray}  \label{2}
\langle \ln t \rangle & = & ik\cos\theta+\frac{ikQ^2_{\nu}\cos3\theta}{%
6\cos^2\theta}- \frac{ikQ^4_{\nu}\sin^2\theta}{2\cos\theta}(1+Q^2_{\nu})
\notag \\
& -& \frac{k^2Q^2_{\nu}\cos^22\theta}{6\cos^2\theta}\left(1+\frac{Q_d^2}{3}%
\right)  \notag \\
& - & \frac{k^2Q^4_{\nu}}{40\cos^2\theta}(2-3\cos2\theta)^2\left (1+\frac{%
Q^2_d}{3}\right)  \notag \\
& - & \frac{k^2Q^6_{\nu}\tan^2\theta}{4}(2-3\cos\theta),  \label{3}
\end{eqnarray}

\begin{eqnarray}
\langle r \rangle & = & -\frac{ikQ^2_{\nu}(2-\cos2\theta)}{6\cos\theta}-
\frac{ikQ^4_{\nu}\sin^2\theta}{2\cos\theta}(1+Q^2_{\nu})  \notag \\
& + & \frac{k^2Q^2_{\nu}}{24\cos^2\theta}(3+10\cos2\theta-\cos4\theta)
\left(1+\frac{Q_d^2}{3}\right)  \notag \\
& - & \frac{k^2Q^4_{\nu}}{80\cos^2\theta}(31-52\cos2\theta+17\cos4\theta)
\left(1+\frac{Q_d^2}{3}\right)  \notag \\
& - & \frac{k^2Q^6_{\nu}\tan^2\theta}{2}(2-3\cos2\theta)+ak^3,  \label{rAvH}
\end{eqnarray}

\begin{eqnarray}
\langle t^2 \rangle & = & 1+2ik\cos\theta-2k^2\cos^2\theta+\frac{%
ikQ^2_{\nu}\cos3\theta}{6\cos^2\theta}  \notag \\
& - & \frac{k^2Q^2_{\nu}}{3\cos^2\theta}(2\cos^2\theta+3\cos^22\theta)
\notag \\
& - & \frac{2k^2Q_d^2}{3}\cos^2\theta,  \label{1}
\end{eqnarray}
where the expression  for the coefficient of the cubic term $a$  in (\ref{rAvH} ) is given in the Appendix.  The transmission length $l_N$, given by Eq. (\ref{FinalN1}), can then be expressed in the asymptotic expansion form as
\begin{eqnarray}
\frac{1}{l_N} & = & \frac{k^2Q^2_{\nu}\cos^22\theta}{6\cos^2\theta}\left(1+\frac{Q_d^2}{3}\right)  \notag \\
& + & \frac{k^2Q_{\nu}^4}{24\cos^4\theta}h_1(\theta)\left (1+\frac{Q_d^2}{3}%
\right )  \notag \\
& + & \frac{k^2Q^6_{\nu}}{17280\cos^6\theta}h_2(\theta)  \notag \\
& + & \frac{Q^4_{\nu}}{18N\cos^4\theta}\left(\frac{3}{2}-\cos^2\theta\right)^2\sin^2(kN\cos\theta),  \label{LHN0}
\end{eqnarray}
where
\begin{eqnarray}
h_1(\theta) & = & 1-\frac{19}{6}\cos2\theta+\frac{7}{15}\cos4\theta + \frac{%
19}{30}\cos6\theta,  \label{lam12HomP} \\
h_2(\theta)& = & 1415-1664\cos2\theta-
188\cos4\theta \nonumber \\
& + & 512\cos6\theta+141\cos8\theta.
\label{h2}
\end{eqnarray}

This expression can be further simplified given $Q_{\nu}\ll1$ and $Q_d\ll1$. We obtain the final form of the transmission length correct to the order of $O(Q^4_{\nu})$:
\begin{eqnarray}
\frac{1}{l_N} & = & \frac{k^2Q^2_{\nu}\cos^22\theta}{6\cos^2\theta}\left(1+%
\frac{Q_d^2}{3}\right) + \frac{k^2Q_{\nu}^4}{24\cos^4\theta}h_1(\theta)
\notag \\
& + & \frac{Q^4_{\nu}}{18N\cos^4\theta}\left(\frac{3}{2}-\cos^2\theta%
\right)^2\sin^2(kN\cos\theta),  \label{LHN}
\end{eqnarray}
with the localization length being given by
\begin{eqnarray}
\frac{1}{l} & = & \frac{k^2Q^2_{\nu}\cos^22\theta}{6\cos^2\theta}+ \frac{%
k^2Q^2_{\nu}Q_d^2\cos^22\theta}{18\cos^2\theta} +  \notag \\
& + & \frac{k^2Q_{\nu}^4}{24\cos^4\theta}h_1(\theta),  \label{LHN2}
\end{eqnarray}
where $h_1(\theta)$ is given by (\ref{lam12HomP}). We have verified numerically the asymptotic formula (\ref{LHN2}) in Sec.\ref{subsec:homo-angle} and found that it is in excellent agreement with the exact numerical calculations for angles of incidence of up to $80^{\circ}$.

Equation (\ref{LHN2}) generalizes the corresponding expression for the localization length obtained in Ref. \cite{PingP} and is applicable to incidence at the Brewster anomaly angle of $\theta=\pi/4$. At this angle, the first two terms in Eq. (\ref{LHN2}) vanish, leading to
\begin{equation}
l=\frac{45\lambda^2}{16\pi^2Q^4_{\nu}}, \ \ \ \theta=\frac{\pi}{4}.
\label{BAA}
\end{equation}
This length is proportional to $Q_{\nu}^{-4}$, in contrast to the $Q_{\nu}^{-2}$ dependence applicable for any incidence angle away from the Brewster angle. However, its wavelength dependence of order $O\left(\lambda^2\right)$ remains the same for all angles less than the critical angle. In the case of weak disorder ($Q_{\nu}\ll 1$), this means that the localization length can be made arbitrarily large at the Brewster anomaly angle relative to that realizable at other incidence angles.

Then, using Eq. (\ref{1}) in Eq. (\ref{crossL}), we may deduce that the characteristic wavelength $\lambda_2$ is identical to that obtained for \emph{s}-polarization (\ref{lam12HomS}). Similarly, $\lambda_1$, obtained using Eq. (\ref{LHN2}), is given by
\begin{equation}
\lambda_1 = \frac{\pi Q_{\nu}}{\cos\theta}\sqrt{\frac{2N}{3}}\sqrt{%
\cos^22\theta+ \frac{Q^2_{\nu}}{\cos^2\theta}h_1(\theta)}.  \label{L1HP}
\end{equation}
The significance of the threshold wavelengths $\lambda_1$ and $\lambda_2$ is the same as for \emph{s}-polarization. Waves with $\lambda<\lambda_1$ are localized, while the wavelength range $\lambda_1<\lambda<\lambda_2$ corresponds to the near ballistic regime. Similarly, the ballistic length $b$ has the same form as the localization length (\ref{BAA})
\begin{equation}
b=\frac{45\lambda^2}{16\pi^2Q^4_{\nu}}, \ \ \ \theta=\frac{\pi}{4}.
\label{BAA-1}
\end{equation}
Thus, the transition from localization to the near ballistic regime takes place without any change of the scale. The wavelength range $\lambda>\lambda_2$ is the far ballistic region in which the ballistic length $b_{f}$ is given by
\begin{eqnarray}
\frac{1}{b_f} & = & \frac{k^2Q^2_{\nu}\cos^22\theta}{6\cos^2\theta}\left(1+%
\frac{Q_d^2}{3}\right)  \notag \\
& + &\frac{Nk^2Q^4_{\nu}}{18\cos^2\theta}\left(\frac{3}{2}%
-\cos^2\theta\right)^2.  \label{LHN3}
\end{eqnarray}
We emphasize that the results obtained in this subsection are applicable only to homogeneous stacks composed of normal material or metamaterial layers.

\subsection{Long-wave asymptotics for mixed stacks}
\label{subsec:long-mix}

\subsubsection{Mixed stacks: $\mathit{s}$-polarization}
\label{subsubsec:MLongE}

Substituting the asymptotic forms (\ref{1E})--(\ref{3E}) into Eqs (\ref{PRB2}) and (\ref{PRB3}), we derive an expression for the reciprocal transmission length:
\begin{eqnarray}
\frac{1}{l_N} & = & \frac{k^2Q^2_{\nu}}{3\cos^2\theta} \left (\frac{1}{2}-%
\frac{1-f(N\alpha_{s})}{3+\zeta\cos^4\theta} \right),  \label{mixE}
\end{eqnarray}
where
\begin{equation}  \label{alpha-s}
\alpha_{s}=\frac {k^2Q^2_{\nu}} {3\cos^2\theta} (3+\zeta\cos^4\theta),
\end{equation}
the function $f$ is as defined in Eq. (\ref{f}), and
\begin{equation}  \label{zeta}
\zeta=\frac {2Q^2_{d}} {Q^2_{\nu}}.
\end{equation}

Eq. (\ref{mixE}) describes the transition from localization to ballistic propagation at long wavelengths and, in the limit as $N\rightarrow\infty$, we obtain the following expression for the localization length
\begin{eqnarray}
l = \frac{3\lambda^{2}\cos^2\theta} {2\pi^2Q^2_{\nu}} \ \frac{3+\zeta\cos^4\theta}{1+\zeta\cos^4\theta}.  \label{mixEL}
\end{eqnarray}
The ballistic length formally corresponds to the opposite extreme, i.e., as $N \to 0$,
\begin{eqnarray}
b = \frac{3\lambda^{2}\cos^2\theta}{2\pi^2Q^2_{\nu}},  \label{mixEB}
\end{eqnarray}
and coincides with the result for a H-stack in \emph{s}-polarization.

The characteristic wavelengths $\lambda_1$ and $\lambda_2$ take the form
\begin{eqnarray}
\lambda_1 & = & \frac{\pi Q_{\nu}}{\cos\theta}\sqrt{\frac{2N}{3}} \sqrt{\frac{1+\zeta\cos^4\theta}{3+\zeta\cos^4\theta}},  \label{mixL1S} \\
\lambda_2 &=& \frac{\pi Q_{\nu}}{\cos\theta} \sqrt{\frac{4N}{3}} \sqrt{3+\zeta\cos^4\theta}.  \label{mixL2S}
\end{eqnarray}
Again, for the range $\lambda\leq \lambda_{1}(N)$, waves are localized, while ballistic propagation occurs for very long wavelengths $\lambda\geq \lambda_{2}(N)$. The transition wavelengths $\lambda_{1,2}$ are of the same order and the intermediate region $\lambda_1(N)<\lambda<\lambda_2(N)$ corresponds to the crossover between localization and ballistic propagation.

\subsubsection{Mixed stacks: $\mathit{p}$-polarization}
\label{subsubsec:MLongH}

While we have derived a general expression for the transmission length that is applicable at arbitrary angles of incidence for a M-stack in \emph{p}-polarization, its form is quite complex and so it is presented only in the Appendix. In what follows, we look at a number of particular cases.

For incidence at angles away from the Brewster angle, the transmission length, according to Eq. (\ref{lpmix}), is given by:
\begin{eqnarray}
\frac{1}{l_N} = \frac {k^2Q^2_{\nu}\cos^{2} 2\theta} {3\cos^2\theta} \left (%
\frac{1}{2}-\frac{1-f(N\alpha_{p})} {2+\cos^{2} 2\theta+\zeta\cos^4\theta}
\right),  \label{lmp}
\end{eqnarray}
where
\begin{equation}  \label{alpha-p}
\alpha_{p}=\frac {k^2Q^2_{\nu}} {3\cos^2\theta} (2+\cos^{2}
2\theta+\zeta\cos^4\theta).
\end{equation}

The localization length may be deduced from Eq. (\ref{lmp}) by taking the limit as $N\rightarrow \infty$, i.e.,
\begin{eqnarray}
{l} = \frac{3\lambda^{2}\cos^2\theta}{2 \pi^2Q^2_{\nu}\cos^2 2\theta} \frac{%
2+\cos^2 2\theta+\zeta\cos^4\theta} {\cos^2 2\theta+\zeta\cos^4\theta}.
\label{lmpL}
\end{eqnarray}

Correspondingly, the ballistic length may be obtained by calculating the limit as $N\rightarrow 0$ in Eq. (\ref{lmp}):
\begin{eqnarray}
{b} = \frac{3\lambda^{2}\cos^2\theta}{2 \pi^2Q^2_{\nu}\cos^2 2\theta}.
\label{lmpb}
\end{eqnarray}

The transmission length for the Brewster anomaly angle can be deduced by substituting $\theta=\pi/4$ in (\ref{lpmix}):
\begin{equation}
\frac{1}{l_N}=\frac{4k^2Q^4_{\nu}}{45} \ \left(1+\frac{121Q_{\nu}^2}{60}
\frac{\left (1-\displaystyle{\frac{5\zeta}{44}}\right )^2}{1+\displaystyle{\frac{\zeta}{8}}}
f(N\alpha_{p})
\right).
\label{tpm4}
\end{equation}
Here, the second term in parentheses is always smaller than the first and therefore, at the Brewster angle, the transmission length as a function of the wavelength exhibits the same dependence, i.e.,
\begin{eqnarray}
l_{N}=l=b=\frac{45\lambda^{2}}{16\pi^{2}Q^4_{\nu}},  \label{tpmLL}
\end{eqnarray}
and is independent of the length of the stack. As in the case of \emph{s}-polarization, the transmission length in the localized regime for \emph{p}-polarization behaves as $O\left( Q^{-4}_{\nu} \right)$ and exceeds the localization length far from the Brewster angle. Note that the localization length (\ref{tpmLL}) for the M-stack at $\theta=\pi/4$ is the same as for
the homogeneous stack (\ref{BAA}).

The transition between the localized and ballistic regimes is again described by two characteristic wavelengths $\lambda_1$ and $\lambda_2$:
\begin{eqnarray}
\lambda_1 & = & \frac{\pi Q_{\nu}\cos2\theta}{\cos\theta} \sqrt{\frac{4N}{3}}
\sqrt{\frac{\cos^2 2\theta+\zeta\cos^4\theta}{2+\cos^2
2\theta+\zeta\cos^4\theta}},  \label{mixL1P} \\
\lambda_2 &=& \frac{\pi Q_{\nu}}{\cos\theta} \sqrt{\frac{4N}{3}} \sqrt{%
4+2\cos^2 2\theta+\zeta\cos^4\theta}.  \label{mixL2P}
\end{eqnarray}
The expression for $\lambda_1$ (\ref{mixL1P}) is obtained under the assumption that the angle of incidence is sufficiently far from the Brewster angle. At the Brewster angle, $\lambda_1$ and $\lambda_2$ are given by
\begin{equation}
\lambda_1=\frac{4\pi Q_{\nu}^2}{3}\sqrt{\frac{N}{5}},  \label{BrA}
\end{equation}
\begin{equation}
\lambda_2=4\pi Q_{\nu} \sqrt{\frac{2N}{3}} \ \sqrt{1+\frac{\zeta}{16}}.
\label{BrB}
\end{equation}
Note that $\lambda_2\gg\lambda_1$. This means that at the Brewster angle, waves such that $\lambda\leq \lambda_{1}(N)$ are localized, while the ballistic region $\lambda_{1}(N)\leq \lambda$ becomes divided into two subregions. The near ballistic subregion is bounded by the two characteristic lengths, i.e., $\lambda_{1}(N)\leq\lambda\leq \lambda_{2}(N)$, and the far ballistic region corresponds to very long waves $\lambda_{2}(N)\leq\lambda.$ As we explained previously, the localization length and the ballistic lengths in each of the two ballistic subregions are described by the same expression (\ref{tpmLL}).

%%%%%%%%%%%%%%%%%%%%%%%%%%%%%%%%%

\section{Results of numerical simulations}
\label{numerics}

We now present results of our comprehensive numerical study of the
properties of the transmission length as a function of wavelength and angle
of incidence. Results are presented for direct simulations based on exact
recurrence relations (\ref{recT})--(\ref{rec}), and are compared with those
obtained from the analytic forms (\ref{PRB1})--(\ref{PRB3}), (\ref{FinalN1}%
), and (\ref{FinalN2}), and their short and long wavelength asymptotic forms
derived in Sec.~\ref{asymp}.

\subsection{Homogeneous stacks}
\label{subsec:homo-numeric}

\subsubsection{Subcritical angle of incidence}
\label{subsubsec:homo-below-critical}

We consider transmission through a H-stack characterized by the parameters: $%
Q_\nu=0.1$, $Q_d=0.2$ at the incidence angle $\theta=45^\circ$, which is
less than the critical angle $\theta_{c}=\sin^{-1}(0.9)\approx 64.16^{\circ}$
and coincides with the Brewster anomaly angle for a layer with a mean
refractive index of $\nu=1$.

\begin{figure}[h]
\center{\rotatebox{0}{\scalebox{0.45}{\includegraphics{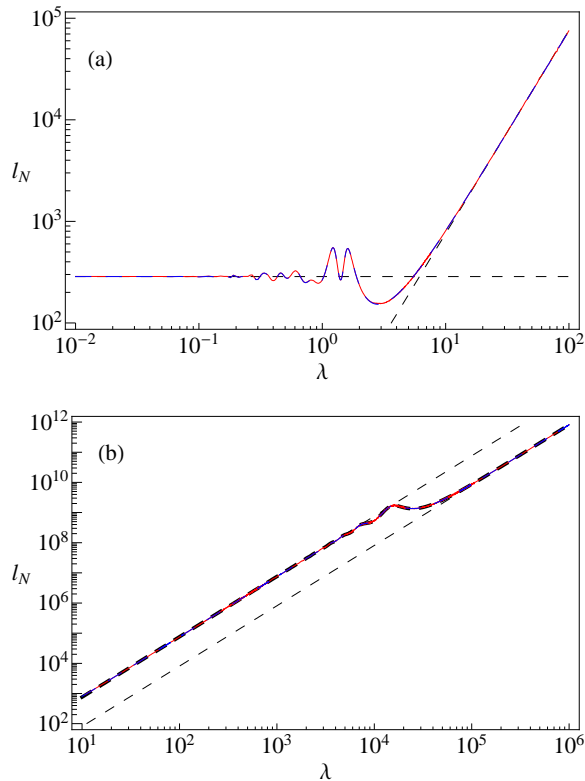}}}}
\caption{(Color online) Transmission length $l_N$ versus wavelength $\protect%
\lambda$ for $Q_{\protect\nu}=0.1$, $Q_d=0.2$ and $\protect\theta=45^\circ$
for \emph{s}-polarized waves; panels (a) localized part of the spectrum and
(b) ballistic part of the spectrum. Red solid curve: numerical simulation;
Blue dash curve analytic form (\protect\ref{FinalN1}). }
\label{Fig2}
\end{figure}

We begin with \emph{s}-polarization, and consider a stack of length $N=10^4,$
using $N_{r}=10^4$ realizations for ensemble averaging. For the given
parameters, the characteristic wavelengths are $\lambda_{1}\approx 36$ and $%
\lambda_{2}\approx 8.9\times 10^{4}$. Plotted in Fig. \ref{Fig2} is the
transmission length as a function of wavelength. Fig. \ref{Fig2}(a)
corresponds to relatively short wavelengths $\lambda\leq 10^{2}$ and
represents mainly the localized part of the spectrum where $N\geq\l %
_{N}\approx l$. Fig. \ref{Fig2}(b) corresponds to longer waves and mostly
displays the ballistic part of the spectrum where $l_{N}\geq N$. In both
panels, the red solid lines represent $l_{N}(\lambda)$, obtained by exact
numerical simulation, and the blue dashed lines display the analytical form (%
\ref{FinalN1}). The excellent agreement between these two curves for all
wavelengths (in both panels) is evident.
%%%%%%%%%%%%%%%%%%%%%%%%%%%%%%%%%%%%%%%%%%%%%%%%%%%%%%%%%%%%%%%%%%%%%%%%%%%%%%%%%%%%%%
\begin{figure}[h]
\center{
\rotatebox{0}{\scalebox{0.45}{\includegraphics{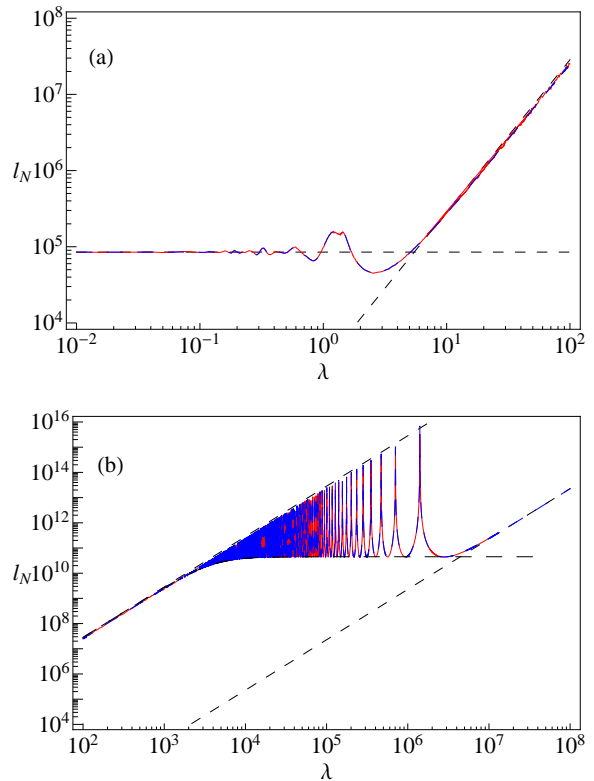}}}}
\caption{(Color online) Transmission length $l_N$ versus $\protect\lambda$
for $Q_{\protect\nu}=0.1$, $Q_d=0.2$ for $p$-polarized waves at the Brewster
anomaly angle $\protect\theta=45^0$. Panel (a) localized part of the
spectrum; panel (b) the transition from localization to ballistic
propagation. Red solid curve: numerical simulation; blue dash curve:
analytic form (\protect\ref{FinalN1}). }
\label{Fig3}
\end{figure}
%%%%%%%%%%%%%%%%%%%%%%%%%%%%%%%%%%%%%%%%%%%%%%%%%%%%%%%%%%%%%%%%%%%%%%%%%%%%%%%%%%%%%%%

The curves displayed in Fig.~\ref{Fig2}(a) explicitly confirm the
coincidence of the long wave asymptotes of the transmission length in both
the localized ($\lambda< 36,$ $l_{N}=l$) and near ballistic ($36<\lambda,$ $%
l_{N}=b$) regions. The slanted, dashed line corresponds to the asymptotic
forms (\ref{LoLE}) and (\ref{balE}) while the horizontal dashed line
corresponds to the short wave asymptote (\ref{shLasEa}). The corresponding
curves of Fig.~\ref{Fig2}(b) display the transmission length in the
ballistic regime. The near ballistic region, where $l_{N}=b$, occurs for $%
36\leq \lambda\leq 8.9\times 10^{4}$, while the transition to the far
ballistic region, where $\l _{N}=b_{f}$, occurs for $\lambda\approx
8.9\times 10^{4}.$ The upper and lower dashed lines respectively display the
near and far ballistic lengths of Eqs (\ref{balE}) and (\ref{ltrLE}). We
observe that the results for \emph{s}-polarization are entirely consistent
with those reported previously for the case of normal incidence \cite%
{Asatryan10}.

Figure~\ref{Fig3} presents the corresponding results for the case of \emph{p}
-polarized waves. Here, we consider a much longer stack of $N=10^6$
layers, the characteristic wavelengths of which are $\lambda_{1}\approx 19$
and $\lambda_{2}\approx 8.9\times 10^{6}$. Fig. {\ref{Fig3}(a) displays the
transmission length spectrum for comparatively short wavelengths $%
\lambda\leq\ 10^{2}$ and corresponds mainly to the localized part of the
spectrum where $N\geq\l _{N}\approx l$. The results of Fig. {\ref{Fig3}(b)
correspond to longer waves $\lambda\geq\ 10^{2}$ and display the ballistic
part of the spectrum $\lambda\geq\lambda_{1}$ where $l_{N}\geq N$. In both
panels, the red solid and the blue dashed lines respectively display $%
l_{N}(\lambda)$ obtained by exact numerical calculation and the analytic
form (\ref{FinalN1}), and we see that their agreement is excellent. }}

The results of Fig.~\ref{Fig3}(a) show that the long wave asymptote of the
transmission length in both the localized region $\lambda<19$ (where $l_{N}=l
$), and the near ballistic region $19<\lambda<10^{2}$ (where $l_{N}=b$)
coincide exactly. The slanted dashed line corresponds to the asymptotic form
(\ref{BAA}) while the horizontal dashed line represents the short wave
asymptote (\ref{ShAs3}). We observe that the short and long wave limits for
the localization length at the Brewster angle are proportional to $%
Q_{\nu}^{-4}$ and hence are two orders larger than in the case of \emph{%
s}-polarization. These numerical results confirm the analytical results
presented earlier in Sec.~\ref{subsubsec:NLongH}.

Fig.~\ref{Fig3}(b) characterizes the ballistic regime which comprises a near
ballistic region ($19\leq \lambda\leq 8.9\times 10^{6}$) in which $l_{N}=b$,
and a far ballistic region where $\l _{N}=b_{f}$, with the transition
between the two occurring at $\lambda\approx 8.9\times 10^{6}$.

The upper and lower dashed lines respectively display the asymptotes for the
near (\ref{BAA-1}) and far (\ref{LHN3}) ballistic lengths. The ballistic
length, over the entire ballistic region, including the transition from the
near to far ballistic regime, is very well described by Eq. (\ref{LHN}). The
oscillatory nature of this transition is due to Fabry-Perot resonances
between the first and the last interfaces of the stack and is much more
pronounced than for the case of \emph{s}-polarization. We observe that the
envelope of the transmission length curve is confined from below by Eq. (\ref%
{LHN}) in which the sine term is replaced by unity. This is the long dashed
black curve of Fig. \ref{Fig3}(b). While this highly oscillatory region also
occurs for \emph{s}-polarized waves, it is not apparent in Fig. %
\ref{Fig2} since the chosen stack length ($N=10^{4}$) was not sufficiently
long to exhibit the feature.

In the samples with only refractive index disorder (i.e., $Q_d=0$), the localization length displays strong oscillations at intermediate wavelengths $0.3<\lambda<2$ for both polarizations. Equation (\ref{FinalN2})
is also in an excellent agreement with the numerical calculations. We also note that the thickness disorder smears out these oscillations, with only few oscillations remaining for $Q_d=0.2$ (see  Figs \ref{Fig2}--\ref{Fig3}).

%%%%%%%%%%%%%%%%%%%%%%%%%%%%%%%%%%%%

\subsubsection{Supercritical angle of incidence}
\label{subsubsec:homo-above-critical}

When the angle of incidence exceeds the critical angle, i.e., $\theta>\theta_c=%
\sin^{-1} (1-Q_{\nu})$, the exponential wave decay can be attributed not
only to Anderson localization but also to attenuation inside the individual
layers.
\begin{figure}[h]
\center{
\rotatebox{0}{\scalebox{0.40}{\includegraphics{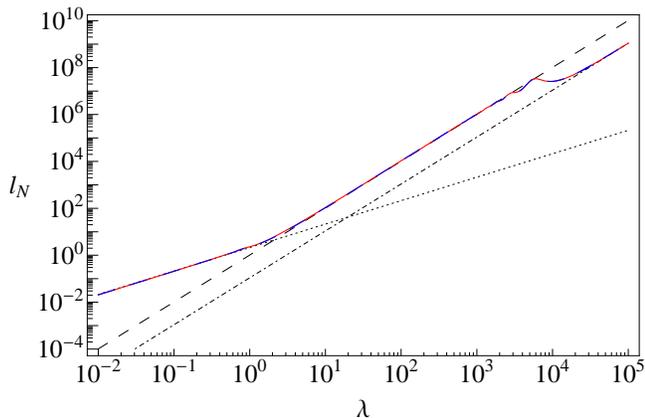}}}}
\caption{(Color online) Transmission length $l_N$ of the homogeneous stack
with $N=10^{4}$, $Q_{\protect\nu}=0.1$, $Q_d=0.2$ versus wavelength $\protect%
\lambda$ for $s$-polarized waves at the supercritical incidence angle $%
\protect\theta=75^\circ$. Red solid curve: numerical simulation; Blue dash
curve analytic form (\protect\ref{FinalN1}). }
\label{Fig6}
\end{figure}

In Fig.~\ref{Fig6}, we plot the transmission length spectrum for a \emph{s}
-polarized wave in which the parameters of the problem are the same as for
Fig.~{\ref{Fig2}, apart from the angle of incidence which is $\theta=75^\circ
$. In this case, the characteristic wavelengths are $\lambda_{1}\approx 99$
and $\lambda_{2}\approx 3.2\times 10^{4}$. Since, as noted previously in
Sec.~\ref{subsec:short},
%we already mentioned at the end of Section \ref{subsec:short}
Eq. (\ref{FinalN1}) can serve as a good interpolation formula for the
transmission length in the short and long wave regions, we have plotted the
results of the exact numerical simulation (red solid curve) together with
those predicted by Eq. (\ref{FinalN1}) (blue dashed curve) to demonstrate
the quality of the agreement. }

In Fig. \ref{Fig6}, the dotted line displays the short wavelength asymptotic
(\ref{shcl}), the black dashed, slanted line displays the localization
length (\ref{LoLE}), and the dashed dotted line shows the far ballistic
asymptotic given by (\ref{ltrLE}). At short wavelengths, $\lambda\leq 2$,
the form of the transmission length spectrum is determined mainly by
attenuation or ``tunneling'' effects. The transmission length is
proportional to the wavelength and is well described asymptotically by Eq. (%
\ref{shcl}), the dotted black slanted line in Fig. \ref{Fig6}. For longer
waves, the form of the transmission length is the same as is observed below
the critical angle, and is described well by Eq. (\ref{ltrE}).

Anderson localization is realized only in the intermediate wavelength
region, $2\leq\lambda\leq 99$, with $l_N \approx l$ with the localization
length given by Eq. (\ref{LoLE}), and shown in the black dashed slanted line
of Fig. ~\ref{Fig6}.

For \emph{p}-polarization, the spectral behavior of the transmission length
is qualitatively equivalent to that for \emph{s}-polarization and so we do
not present this here. There is excellent agreement between the exact
numerical calculation, the theoretical result of Eq. (\ref{FinalN2}), and
the short (\ref{ShAs3}) and long (\ref{LHN2}) wave asymptotic forms.

\subsection{Mixed stacks}
\label{subsec:mix-numeric}

\subsubsection{Subcritical angle of incidence}
\label{mix-below-critical}

We first consider the case of \emph{s}-polarized wave propagation through a
mixed (i.e., alternating layers of normal and meta-materials) stack of length $%
N=10^4$. The parameters are the same as those adopted in Sec. \ref%
{subsubsec:homo-below-critical}, i.e., $Q_\nu=0.1$, $Q_d=0.2$, $N_{r}=10^4$,
and the incidence angle is $\theta=45^\circ$, which is less than the
critical angle $\theta_{c}=\sin^{-1}(0.9) = 64.16^\circ$ and coincides with
the Brewster angle for the single layer with mean refractive index $\nu=\pm 1
$. For these parameters, the characteristic wavelengths given by Eqs (\ref%
{mixL1S}) and (\ref{mixL2S}) are $\lambda_{1}\approx 28$, and $%
\lambda_{2}\approx 115$ correspondingly.

\begin{figure}[h]
\center{
\rotatebox{0}{\scalebox{0.4}{\includegraphics{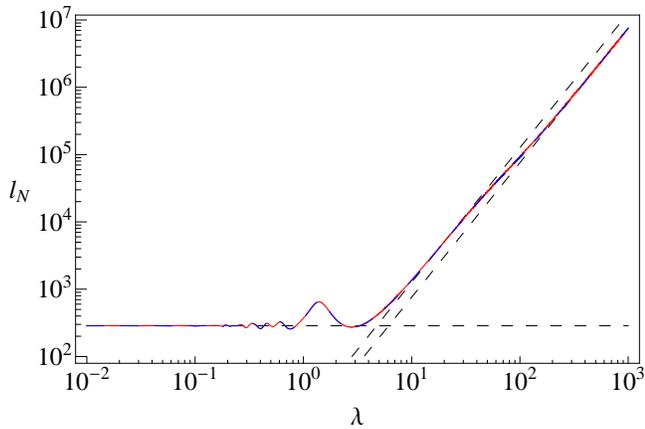}}}}
%\center{\rotatebox{-90}{\scalebox{0.35}{\includegraphics{Fig3off.eps}}}}
\caption{(Color online) Transmission length versus $\protect\lambda$ for a
M-stack in $s$-polarized light with $Q_{\protect\nu}=0.1$, $Q_d=0.2$ and $%
N=10^{4}$ for $\protect\theta=45^{\circ}$. Red solid curve: numerical
simulations; Blue dash curve: analytic form (\protect\ref{PRB1}). }
\label{Fig4}
\end{figure}

In Fig. \ref{Fig4}, the red solid line and the blue dashed line respectively
display results from the numerical simulation and the analytic form (based
on the WSA) for the transmission length as a function of wavelength, with
these two coinciding to high accuracy.

%%%%%%%%%%%%%%%%%%%%%%%%%%%%%%%%%%%%%%%%%%%%%%%%%%%%%%%%%%%%%%%%%%%%%%%%%%%%%%%%%%%%%%
%%%%%%%%%%%%%%%%%%%%%%%%%%%%%%%%%%%%%%%%%%%%%%%%%%%%%%%%%%%%%%%%%%%%%%%%%%%%%%%%%%%%%%%

The form of the transmission length spectrum is similar to that observed for
the case of normal incidence \cite{Asatryan10}. The short wavelength
asymptotic form (\ref{ShAs3}), shown as the horizontal dashed line in Fig. %
\ref{Fig4}, is the same as for a H-stack. For $\lambda\leq\lambda_1=28$, all
waves are localized, with the localization length (\ref{mixEL}) shown by the
upper dashed, slanted straight line. The transition from localization to
ballistic propagation occurs in the wavelength range $\lambda_1<\lambda<%
\lambda_2$ and is well described by Eq. (\ref{mixE}). Ballistic propagation
occurs for $\lambda\geq\lambda_2=115$ and is characterized by the ballistic
length (\ref{mixEB}) which differs from the M-stack localization length (\ref%
{mixE}), in contrast to the case of H-stacks.

\begin{figure}[h]
\center{
\rotatebox{0}{\scalebox{0.4}{\includegraphics{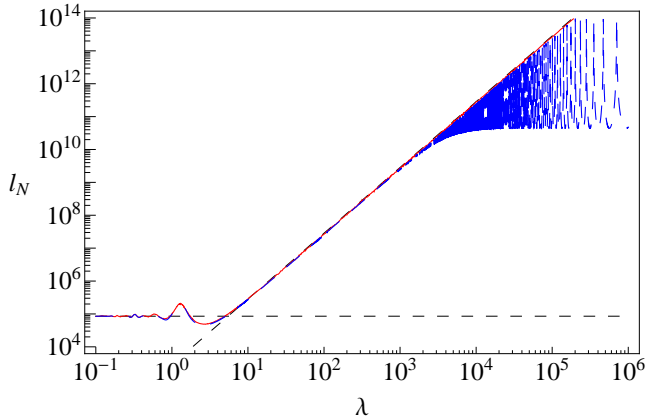}}}}
\caption{(Color online) Transmission length versus $\protect\lambda$ for a
M- stack in $p$-polarized light with $Q_{\protect\nu}=0.1$, $Q_d=0.2$ and $%
N=10^{6}$, at the Brewster angle $\protect\theta=45^{0}$ red solid line. The
blue dashed line shows results for \emph{s}-polarization and a H-stack,
re-plotted for comparison. }
\label{Fig5}
\end{figure}

Figure \ref{Fig5} displays the transmission length spectrum for a M-stack of
length $N=10^6$ in \emph{p}-polarized light, with all other parameters
identical to that for the \emph{s}-polarization simulations. In this case,
the characteristic wavelengths are $\lambda_1\approx 19$ (from Eq.(\ref%
{mixL1P})) and $\lambda_{2} \approx 1200\pi$ (from Eq.(\ref{mixL2P})).

The results of the numerical simulation and the WSA analytical forms (\ref%
{PRB1}), (\ref{PRB2}) coincide and are displayed by a single red solid line.
Localization occurs for $\lambda\leq\lambda_1\approx 19$, while the
transition from localization to ballistic propagation occurs at $%
\lambda\sim\lambda_1$. In contrast to the case of \emph{s}-polarization,
the transition is not accompanied by a change of scale and is given by the
same wavelength dependence (\ref{tpmLL}). The same asymptotic (\ref{tpmLL})
also holds for wavelengths $\lambda>\lambda_{2}\approx 1200\pi$, which
defines the transition from the near to the ballistic regime. As a
consequence of the disorder $Q_{\nu}=0.1$, the short wave localization
length (\ref{ShAs3}) (horizontal dashed line) is two orders of magnitude
larger than that for \emph{s}-polarized light (cf. Fig.\ref{Fig4}).

\subsubsection{Supercritical angle of incidence}
\label{subsubsec:mix-above-critical}

\begin{figure}[h]
\center{
\rotatebox{0}{\scalebox{0.4}{\includegraphics{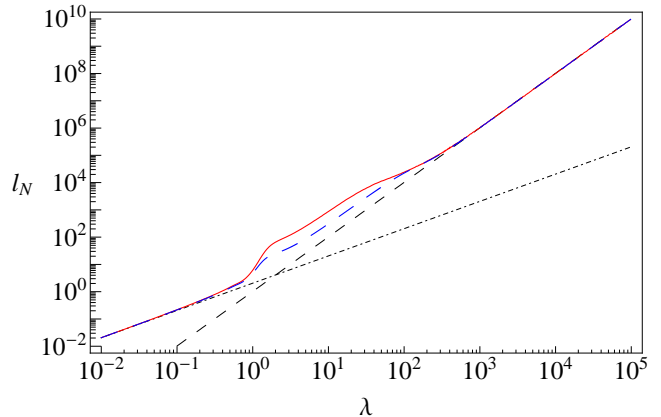}}}}
\caption{(Color online) Transmission length versus $\protect\lambda$ for a
M-stack in $s$-polarized light with $Q_{\protect\nu}=0.1$, $Q_d=0.2$ and $%
N=10^{4}$, and for the supercritical incidence angle $\protect\theta%
=75^{\circ}$. Red solid curve: numerical simulations; Blue dash curve:
analytic form (\protect\ref{PRB1}). }
\label{Fig8}
\end{figure}

We now consider a case in which the angle of incidence $\theta=75^{\circ}$
exceeds the critical angle $\theta_c=\sin^{-1}(1-Q_{\nu})=64.16^{\circ}$ (%
\ref{theta-c}). In Fig. \ref{Fig8} we present the transmission length
spectrum for \emph{s}-polarized light and display results from
the exact numerical calculation (red solid line) and the analytic form (long
dashed blue curve). The expectation that Eq. (\ref{PRB1}) would serve as a
good interpolation formula for the transmission length in the short and long
wave regions, as anticipated in Sec. \ref{subsec:short}, is borne out by the
results of Fig. \ref{Fig8}. The short wave (dashed dotted line) asymptotic
form (\ref{shcl}) and the long wave (black dashed line) asymptotic form (\ref%
{mixEB}) respectively coincide with the numerical results for $\lambda\leq 1$
and $200\leq\lambda$. In the intermediate region $1\leq\lambda\leq 200$,
however, the theoretical description underestimates the actual transmission
length since the WSA is no longer valid for the chosen, supercritical angle
of incidence. For \emph{p}-polarization, the results are qualitatively the
same, but with the discrepancy at the intermediate wavelengths even more
pronounced.

\subsection{Mixed stacks with refractive-index disorder}
\label{sec:only-refrac}

In our earlier paper \cite{Asatryan07}, we demonstrated that at normal
incidence a disordered mixed stack, with only refractive index disorder,
could substantially suppress Anderson localization. Indeed, the suppression
is so strong that even the usual quadratic dependence on wavelength (i.e., $%
O(\lambda^2)$) of the localization length at long wavelengths was shown to
change to $O(\lambda^6)$. In contrast, the introduction of the thickness
disorder in combination with the refractive index disorder induces strong
localization at long wavelengths, with the localization length returning to
its expected quadratic dependence on wavelength \cite{Asatryan10}. In this
section, we consider the effects of polarization on long wavelength
localization in M-stacks.

\begin{figure}[h]
\center{
\rotatebox{0}{\scalebox{0.4}{\includegraphics{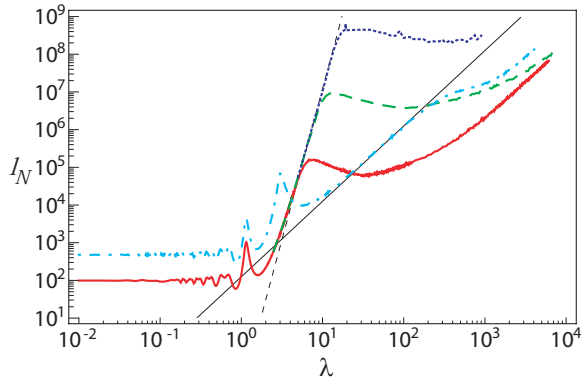}}}}
\caption{(Color online) Transmission length $l_N$ versus $\protect\lambda$
for a M-stack with $Q_{\protect\nu}=0.25$, $Q_d=0$ and $\protect\theta=30^0$
for \emph{p}-polarized light (cyan dashed dotted curve, $N=10^6$) and \emph{%
s}-polarized light (red solid curve, $N=10^5$; green dashed curve, $N=10^7$;
blue dotted curve, $N=8\times10^8$). }
\label{Fig10A}
\end{figure}

Figure \ref{Fig10A} displays transmission length spectra for a mixed stack
with only refractive index disorder for an angle of incidence of $%
\theta=30^\circ$. Four curves are displayed: for \emph{p}-polarized light
and a stack of length $N=10^{6}$ (dashed doted cyan curve), and for \emph{%
s}-polarized light and three stacks of lengths $N=10^5$ (solid red curve), $%
N=10^7$ (dashed green curve) and $N=8\times 10^8$ (blue curve). There is a
striking difference between the two polarizations: in the case of \emph{%
p}-polarized light, there is strong localization at long wavelengths ($%
\lambda\leq10^2$), with the localization length showing $O(\lambda^2)$
dependence; in contrast, the localization length for \emph{s}-polarized
light is much larger and shows the $O(\lambda^6)$ dependence as occurs for
normal incidence. Note that for \emph{s}-polarization, the localization
regions in Fig. \ref{Fig10A} are bounded from above by the wavelength limits
$\lambda\leq5,9$, and $12$ for stacks of length $N=10^5,10^{7}$, and $8
\times 10^{8}$ respectively.

This asymmetry between the polarizations suggests that the suppression of
localization is due not only to the suppression of the phase accumulation
but also to the vector nature of the electromagnetic wave. Because of the
symmetry of Maxwell's equations between the electric and magnetic fields, it
is to be expected that for a model in which there is disorder in the
magnetic permeability (with $\varepsilon=\pm1$) the situation will be
inverted, with localization for \emph{p}-polarized waves being
suppressed and with \emph{s}-polarization showing strong localization.

In concluding this section, we emphasize that the delicate phenomenon of the suppression of localization occurs only for refractive index disorder, and that the introduction of any thickness disorder leads to the strong localization (see Sec.~\ref{mix-below-critical} and Ref.\cite{Asatryan10}).

\section{Transmission length as a function of the incidence angle}
\label{theta}

\subsection{Homogeneous stacks}
\label{subsec:homo-angle}

\begin{figure}[h]
\center{
\rotatebox{0}{\scalebox{0.45}{\includegraphics{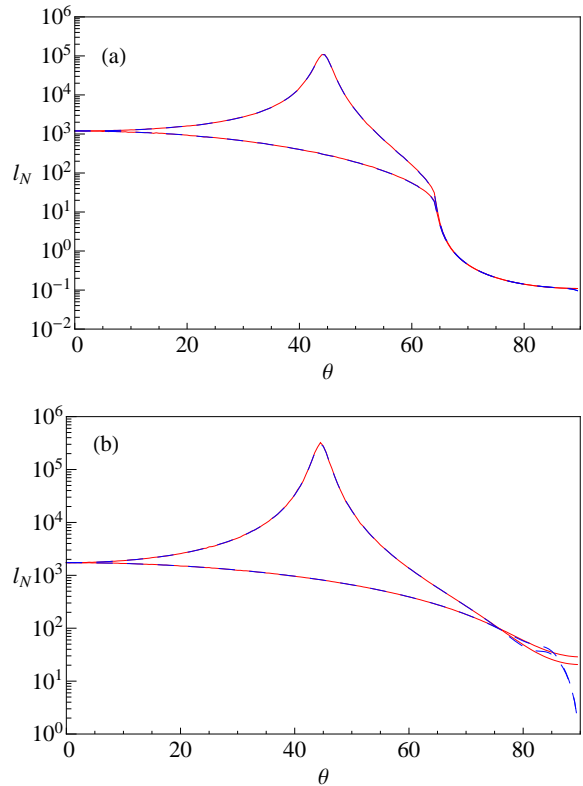}}}}
\caption{(Color online) Transmission length $l_N$ versus incidence angle $%
\protect\theta$ for a homogeneous stack with $Q_{\protect\nu}=0.1$, $Q_d=0.2$
for (a) $\protect\lambda=0.1$ (upper panel), (b) $\protect\lambda=10$ (lower
panel). Red solid curve: numerical simulations; Blue dash curve analytic
form (\protect\ref{FinalN2}). The top set of curves in each of the panels
are for \emph{p}-polarization while the bottom set of curves are for \emph{%
s}-polarization. }
\label{Fig11}
\end{figure}

We next consider the angular dependence of the transmission length of a
homogeneous stack for a given wavelength. As in earlier simulations, we work
with the parameters $Q_\nu=0.1$, $Q_d=0.2$, and $N=10^6$. Figure~\ref{Fig11}
displays the transmission length as a function of the angle of incidence $%
\theta$ for both \emph{s}- and \emph{p}-polarizations. In each panel
(upper: $\lambda=0.1$, lower: $\lambda=10$), the solid red curve displays
the results of the numerical simulation while the blue dashed line
corresponds to the WSA analytic form (\ref{FinalN2}), with the top and
bottom sets being for \emph{p}- and \emph{s}-polarizations respectively.

For the short wavelength $\lambda=0.1$ (Fig. \ref{Fig11}(a)), the analytic
form agrees perfectly with the simulations. While for \emph{s}-polarization,
the transmission length decreases monotonically with the angle of incidence,
the transmission length for \emph{p}-polarization displays a pronounced
maximum at the Brewster anomaly angle (at $\theta\approx46^\circ$ for these
parameters). In the supercritical regime, $\theta>\theta_{c} \approx
64^{\circ}$, attenuation is the dominant mechanism for localization and
hence the behavior of the two polarizations coincide.

For long wavelengths, as in Fig. \ref{Fig11}(b), we see that for extreme
angles of incidence (e.g., for $\theta>80^\circ$ for the wavelength $%
\lambda=10$), the theoretical prediction departs markedly from the
simulation results. Similar departures for intermediate wavelengths (e.g,
for $\lambda=1$) also exist for angles of incidence $\theta>85^\circ$.

We have also calculated the localization length for the very long wavelength of $\lambda=40$ as a function of the angle of incidence (for the same parameters as for Fig.\ref{Fig11}), for which the expansion (\ref{LHN2}) is applicable. There is excellent agreement between the exact numerical calculation and the
asymptotic form (\ref{LHN2}) for angles of up to $80^{\circ}$.  (Since this plot is very similar to Fig.\ref{Fig11}(b), it is not included in the manuscript.)

\subsection{Mixed stacks}
\label{subsec:mix-angle}

\begin{figure}[h]
\center{
\rotatebox{0}{\scalebox{0.45}{\includegraphics{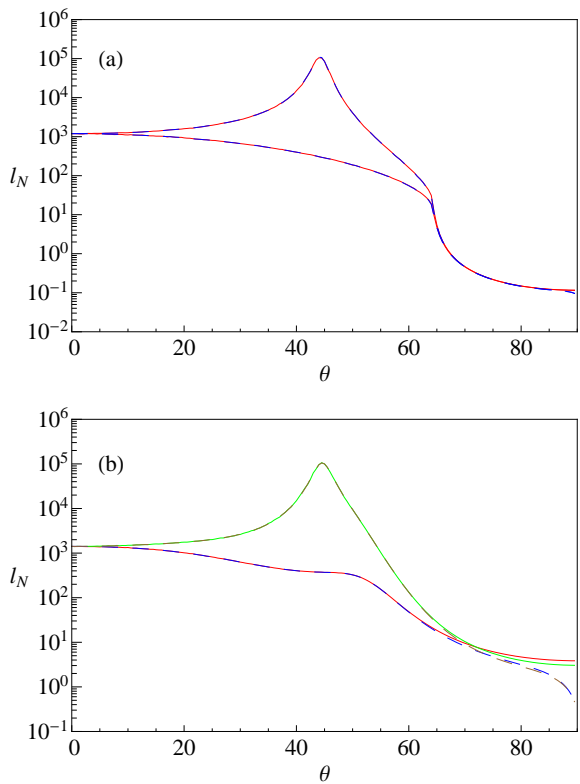}}}}
\caption{(Color online) Transmission length $l_N$ versus incidence angle $%
\protect\theta$ for a mixed stack with $Q_{\protect\nu}=0.1$, $Q_d=0.2$, for
(a) $\protect\lambda=0.1$ (upper panel), and (b) $\protect\lambda=1$ (lower
panel). The top and bottom curves are respectively for \emph{p-} and \emph{s}-%
polarizations. }
\label{Fig22}
\end{figure}

We now consider the angular dependence of the transmission length for mixed
stacks and, in Fig.~\ref{Fig22}, we plot $l_{N}$ as a function of the angle $%
\theta$ for a stack of length $N=10^{6}$ at the two wavelengths $\lambda=0.1$
(Fig. \ref{Fig22}(a)) and $\lambda=1$ (Fig. \ref{Fig22}(b)). In either case,
the calculated transmission length does not exceed the stack length and so,
for subcritical angles, our calculations display the true localization
length. For the shorter wavelength $\lambda=0.1$, the form of the
transmission length for both polarizations is similar to that observed for
homogeneous stacks (cf. Fig. \ref{Fig11}(a)), and we also note that the
analytical form (\ref{PRB2}) agrees perfectly with the results from the
numerical simulations.

Fig. \ref{Fig22}(b) displays results for an intermediate wavelength $%
\lambda=1$ with the lower solid red and blue dashed curves respectively
displaying the results of numerical simulations and analytical predictions (%
\ref{PRB2}) for \emph{s}-polarization, (bottom curves), while the upper
solid green and brown dashed curves display simulations and analytical
predictions (\ref{PRB2}) for \emph{p}-polarization. The agreement between
simulations and the theoretical form is again excellent for angles of
incidence less then the critical angle, $\theta<\theta_{c}$, while for
angles greater then the critical angle, the discrepancies that are evident
are again explicable by the breaking down of the WSA at extreme angles of
incidence.

\subsection{Alternating homogeneous stacks}
\label{subsec:alt-homo-angle}

In this section, we present an example of true delocalization arising from
the vector nature of the electromagnetic field. This was first pointed out
by Sipe \textit{et al} \cite{Sipe}, in which an analysis applicable at long
wavelengths was presented. More recently, the analysis has been extended to
short wavelengths \cite{Bliokh}. The condition for the Brewster anomaly can
be satisfied for a homogeneous stack (i.e., with all layers being either
normal materials or all being metamaterials) with only thickness disorder,
and with alternating refractive indices (i.e., with refractive indices $%
\nu_{A}$ and $\nu_{B}$ respectively for odd and even numbered layers).

\begin{figure}[h]
\center{
\rotatebox{0}{\scalebox{0.45}{\includegraphics{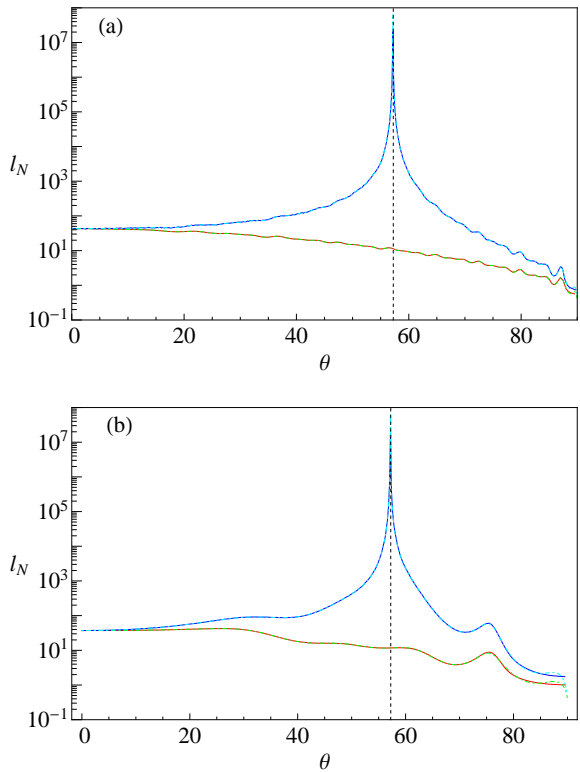}}}}
\caption{ (Color online) Delocalization at the Brewster angle $\protect\theta%
_B$; (a) at a short wavelength $\protect\lambda=0.1$, and (b) an
intermediate wavelength $\protect\lambda=0.5$. Top and bottom curves are
respectively for \emph{p-} and the \emph{s-}polarizations. }
\label{Fig13}
\end{figure}

We proceed in a similar manner to that of Ref. \cite{Sipe} and consider a
stack in vacuum with $\nu_{A}=1$ and $\nu_{B}=1.5$, and with layers whose
random thicknesses are uniformly distributed in the interval $d \in [0.8,
1.2]$ (i.e, $Q_d=0.2$). We note that in the case of \emph{p}-polarization,
the applicability of the weak scattering approximation is heightened in the
vicinity of the Brewster angle since each layer is almost transparent.

In Figs \ref{Fig13} (a) and (b), we respectively plot the transmission
length as a function of the angle of incidence at a short wavelength $%
\lambda=0.1$, and also at an intermediate wavelength $\lambda=0.5$. The
lower and upper curves are respectively for \emph{s-} and  \emph{p-}%
polarizations, and we see, somewhat surprisingly, that the numerical
simulations and the analytic forms obtained within a WSA framework are
essentially identical for both polarizations for arbitrary incidence angles,
apart from the discrepancies evident for extreme angles $\theta>87^\circ$ at
$\lambda=0.5$. The surprising element is that the theoretical framework
based on the WSA appears to work over a much wider range of angles and
polarizations than that suggested by strict validity of the WSA. In the
figure, the theoretical description for \emph{s}-polarization (green curve)
overlays the results of the numerical calculation (red curve). The same is
true for \emph{p}-polarization, with the theoretical prediction (black
dotted line) overlaying results from the numerical calculation (cyan solid
curve).

In these calculations, the stack length was $N=10^4$ and so waves are
delocalized for incidence angles $55^0\leq\theta\leq 59^0$ around the
Brewster angle $\theta_B=\arctan(1.5)\approx 56.19^\circ$ where the
transmission length $l_N\geq N$. The localization properties for the
corresponding homogeneous stack composed of metamaterial layers with $%
\nu_B=-1.5$ is the same as that shown in Figs \ref{Fig13} for normal layers
with $\nu_B=1.5$. The WSA based theory also appears to work over a
reasonably broad range of wavelengths, although for the intermediate
wavelengths $10\leq\lambda\leq 50$ there are some differences between
simulations and the theoretical prediction.

\section{Conclusions}

\label{conc} We have investigated the effect of polarization on the Anderson
localization in one-dimensional disordered stacks composed entirely of either
right- or left-handed layers, as well as mixed stacks with alternating
sequence of normal and metamaterial layers.

Our analysis has generalized the results obtained earlier \cite%
{Asatryan10} for the case of normal incidence to the case of an arbitrary angle
of incidence, with a particular attention paid to the localization at the Brewster
angle. Based on this approach, we have carried out a comprehensive
study of the localization length as a function of both the angle of
incidence and the polarization of the incident wave for various types of
disorder.

In the case of general disorder, where both refractive index and
thickness of the layers are random, we have derived the long- and short-wave
asymptotics for the localization length for a wide range of incidence
angles, including the Brewster angle. At the Brewster angle, we have shown
that the localization length continues to exhibit a quadratic dependence on
wavelength (as in the case of the normal incidence), but that the coefficient of
proportionality becomes parametrically larger, being proportional to $Q^{-4}$%
, rather than $Q^{-2}$ ($Q\ll 1$), as for the case of the normal incidence.

Our theoretical study not only characterizes the localization and
ballistic propagation, but also describes perfectly the crossover between
these two regimes. We have also shown that the transition from localization
to ballistic propagation in the vicinity of the Brewster angle in a mixed
stack is given by a single scale (\ref{tpmLL}). In the case of thickness
disorder, we have shown that, at the Brewster angle, Anderson localization
is suppressed completely.

%%%%%%%%%%%%%%%%%%%%%%%%%%%%%%%%%%%%%%%%%%%%%%%%%%%%%%%%%%%%%%%%%%%%%%%%

\section{Acknowledgments}
\label{sec:ackn}

This work was supported by the Australian Research Council through its
Discovery Grants program. We also acknowledge the provision of computing
facilities through the National Computational Infrastructure in Australia.

%%%%%%%%%%%%%%%%%%%%%%%%%%%%%%%%%%%%%%%%%%%%%%%%%%%%%%%%%%%%%%%%%%%%%
\appendix*

\section{Transmission length for $p$-polarization and mixed stack}

In this Appendix, we provide an asymptotic form for the transmission length in the case of {\em p}-polarization for a mixed stack. The reciprocal transmission length (\ref{PRB1}) takes the form
\begin{eqnarray}
\frac{1}{l_N} & = & \frac{k^2Q^2_{\nu}}{6\cos^2\theta} \cos^22\theta \left (1+\frac{Q^2_d}{3} \right ) \nonumber \\
& + & \frac{k^2Q^4_{\nu}}{40\cos^2\theta}\left(2-
3\cos2\theta\right)^2\left ( 1+\frac{Q^2_d}{3}  \right ) \nonumber\\
& + & \frac{k^2Q^6_{\nu}}{4}(2-3\cos2\theta)\left (1+\frac{Q^2_d}{3}\right )\tan^2\theta\nonumber\\
& + & \frac{k^2Q^8_{\nu}}{48}(43-55\cos2\theta)\tan^2\theta\nonumber\\
& -&
\frac{\displaystyle{\frac{2k^2Q^4_{\nu}\cos^22\theta}
{3\cos^2\theta}+f_1+
f_2+f_3+f_4+f_5}}{2Q^2_{\nu}(2+\cos^22\theta)+
4Q_d^2\cos^4\theta}\nonumber\\
& + &
\frac{\frac{4k^2Q^4_{\nu}\cos^22\theta}{3\cos^2\theta}+g_1+g_2+g_3+g_4+g_5}
{4Q^2_{\nu}(2+\cos^22\theta)+
8Q_d^2\cos^4\theta}f(N\alpha_p),\nonumber\\
\label{lpmix}
\end{eqnarray}
where
\begin{eqnarray}
f_1 &= & \frac{k^2Q^4_{\nu}Q_d^2}{8\cos^2\theta}\left (\frac{61}{8}+\frac{25}{6}\cos2\theta
 + \frac{19}{3}\cos4\theta \right.\nonumber\\
 & - &  \left.
\frac{5}{6}\cos6\theta
 +  \frac{\cos8\theta}{24}\right ),
\label{f1}\\
f_2 &= & \frac{k^2Q^6_{\nu}}{12\cos^2\theta}\left (\frac{1927}{120}-\frac{121}{5}\cos2\theta+\frac{403}{30}\cos4\theta  \right.\nonumber\\
& - & \left. \frac{49}{15}\cos6\theta  +  \frac{\cos8\theta}{24} \right ),
\label{f2}\\
f_3 &= & \frac{k^2Q^6_{\nu}Q_d^2}{32\cos^2\theta}\left (\frac{7013}{270}-\frac{1763}{45}\cos2\theta+\frac{3772}{135}\cos4\theta  \right.\nonumber\\
& - & \left. \frac{1391}{135}\cos6\theta  +  \frac{163}{270}\cos8\theta \right ),
\label{f3}\\
f_4 &= & \frac{k^2Q^8_{\nu}}{32\cos^2\theta}\left (\frac{633}{4}-\frac{19058}{75}\cos2\theta+\frac{9521}{75}\cos4\theta  \right.\nonumber\\
& - & \left. \frac{818}{25}\cos6\theta  +  \frac{593}{300}\cos8\theta \right ),
\label{f4}\\
f_5 &= & \frac{k^2Q^4_{\nu}Q^4_d}{2592\cos^2\theta}\left (3+10\cos2\theta-\cos4\theta  \right )^2,
\label{f5}
\end{eqnarray}
and
\begin{eqnarray}
g_1 & = & \frac{2k^2Q^6_{\nu}}{45\cos^2\theta}\left ( 41-77\cos2\theta+41\cos4\theta  \right.\nonumber\\
& - & \left. 11\cos6\theta \right ),
\label{g1}\\
g_2 & = & \frac{k^2Q^4_{\nu}Q_d^2}{18\cos^2\theta}\left ( 10 +5\cos2\theta+10\cos4\theta  \right.\nonumber\\
& - & \left. \cos6\theta \right ),
\label{g2}\\
g_3 &= & \frac{k^2Q^8_{\nu}}{270\cos^2\theta}\left (1963-\frac{16534}{5}\cos2\theta+\frac{8968}{5}\cos4\theta  \right.\nonumber\\
& - & \left. \frac{2362}{5}\cos6\theta  +  \frac{121}{5}\cos8\theta \right ),
\label{g3}\\
g_4 &= & \frac{k^2Q^6_{\nu}Q^2_d}{60\cos^2\theta}\left (\frac{247}{3}-139\cos2\theta+\frac{928}{9}\cos4\theta  \right.\nonumber\\
& - & \left. \frac{103}{3}\cos6\theta  +  \frac{11}{9}\cos8\theta \right ),
\label{g4}\\
g_5 &= & \frac{k^2Q^4_{\nu}Q^4_d}{432\cos^2\theta}\left (3+10\cos2\theta-\cos4\theta \right )^2
\label{g5}.
\end{eqnarray}

The expansion (\ref{lpmix}) is valid for any angle of incidence for mixed stacks  at long wavelengths. The  factor $f$ is given by (\ref{f}) in this expression and characterizes the transition from  localization to  ballistic propagation.

For the sake of completeness, we also provide in this Appendix the expression for the cubic coefficient $a$ in the expansion (\ref{rAvH}).
\begin{eqnarray}  \label{2A}
 a  & = & (1+Q^2_{d})\left [-\frac{iQ^2_{\nu}}{36\cos\theta}(3+18\cos2\theta-\cos4\theta)\right.
  \notag \\
& + & \left. \frac{iQ^4_{\nu}}{480\cos^3\theta}(146-87\cos2\theta+158\cos4\theta-41\cos6\theta)
\right.   \notag \\
& - & \left. \frac{iQ^6_{\nu}}{336\cos^3\theta}(365-604\cos2\theta+331\cos4\theta-86\cos6\theta)
\right ]   \notag \\
& - &  \frac{iQ^8_{\nu}\tan^2\theta}{48\cos\theta}(175-264\cos2\theta+107\cos4\theta).  \label{a3}
\end{eqnarray}

All  expansions in (\ref{1E})--(\ref{3E}), (\ref{3})--(\ref{1}) and in the Appendix can be readily obtained by using symbolic manipulation package such as {\em Mathematica}.

%%%%%%%%%%%%%%%%%%%%%%%%%%%%%%%%%%%%%%%%%%

\end{document}